# Micro-displacement sensors based on plastic photonic bandgap Bragg fibers

**H. Qu,[1] T. Brastaviceanu,[2] F. Bergeron,[2] J. Olesik,[2] M. Skorobogatiy[1,*]**

[1]Genie Physique, Ecole Polytechnique de Montreal, C. P. 6079. Succ. Centre-ville, Montreal, Quebec, Canada, H3C 3A7
[2]Tactus Scientific Inc., Campus des Technologies de la Sante, 5795 de Gaspe, local 216, Montreal, Quebec, Canada, H2S 2X3
*Corresponding author: maksim.skorobogatiy@polymtl.ca

We demonstrate an amplitude-based micro-displacement sensor that uses a plastic photonic bandgap Bragg fiber with one end coated with a silver layer. The reflection intensity of the Bragg fiber is characterized in response to different displacements (or bending curvatures). We note that the Bragg reflector of the fiber acts as an efficient mode stripper for the wavelengths near the edge of the fiber bandgap, which makes the sensor extremely sensitive to bending or displacements at these wavelengths. Besides, by comparison of the Bragg fiber sensor to a sensor based on a regular multimode fiber with similar outer diameter and length, we find that the Bragg fiber sensor is more sensitive to bending due to presence of mode stripper in the form of the multilayer reflector. Experimental results show that the minimum detection limit of the Bragg fiber sensor can be smaller than 5 μm for displacement sensing.

In the past several decades, development of low-cost and compact fiber-optic bending / displacement sensors has received significant attention due to their potential applications in various scientific and industrial fields such as bio- and physiological sensing, architecture, robotics, astronautics and vehicle industry, to name a few. According to their sensing mechanism, fiber-optic bending sensors are generally categorized into two classes: intensity-based sensors and spectral-based sensors. Intensity-based sensors use either single-mode or multimode fibers; transmitted intensity through such fibers is then characterized as a function of the bending curvature. Among advantages of this type of bending sensors are low cost, ease of fabrication, and simple signal acquisition and processing since no spectral manipulations are required. However, the detection accuracy of the amplitude-based sensors is prone to errors due to intensity fluctuations of the light source. Moreover, to increase sensor sensitivity, it is preferable to strip as much as possible the cladding modes, which are excited by the bend. This is typically accomplished by coating the fiber cladding with an absorbing layer or by decorating the fiber cladding with scatterers such as scratches [1]. Spectral-based fiber-optic bending sensors normally employ in-fiber resonant structures such as fiber Bragg gratings [2], long period gratings [3], and fiber interferometric structures [4]. Changes in the positions of resonant features in the fiber transmission or reflection spectra are then used to extract bending curvature. Note that fiber sensors based on spectral detection usually require sophisticated and expensive fiber components, and they also require expensive characterization equipment (spectrometers) which are difficult to miniaturize.

In this paper, we demonstrate an amplitude-based fiber bending / displacement sensor that uses plastic photonic bandgap Bragg fiber with one end coated with a silver layer. This type of fiber features a relatively low-loss plastic core surrounded by a multilayer dielectric reflector. This reflector is an efficient mode stripper for the wavelengths outside of the fiber bandgap. In fact, outside of the fiber bandgap, the core guided light penetrates strongly into the multilayer reflector region, which acts as a strong scatterer due to imperfections on the multilayer interfaces. Therefore, by interrogating the fiber at the wavelengths near the bandgap edge, high sensitivities to bending can be achieved. For comparison, we also test bending sensor based on a regular multimode fiber of similar outside diameter and length. We find that Bragg fiber-based sensor is more sensitive to bending as compared to the sensors based on regular multimode fibers due to presence of an efficient mode stripper in the form of a multilayer reflector. Finally, we show that the Bragg fiber bend sensor can be packaged into a compact sensor module by adopting the "MOSQUITO" sensor configuration detailed in [5].

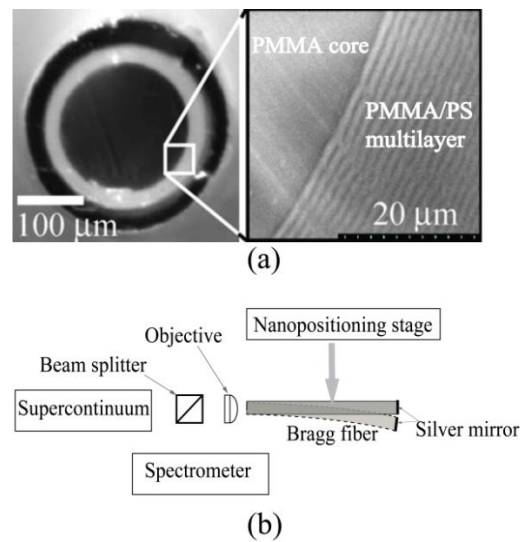

Fig.1 (a) Cross section of a solid-core Bragg fiber; (b) Schematic of the Bragg fiber bending sensor.

The fiber used in this work is an all-polymer solid-core photonic bandgap Bragg fiber fabricated in our group. As shown in Fig. 1 (a), a Bragg fiber features a polymethyl methacrylate (PMMA) core surrounded by a Bragg reflector consisting of an alternating PMMA/polystyrene (PS) multilayer (refractive index: 1.48/1.59). Such a multilayer is responsible for the appearance of a spectrally narrow transmission band (reflector bandgap) within which the light is strongly confined inside of the fiber core, with a minimal presence in the reflector region. For the wavelengths outside of the reflector bandgap, the light penetrates deeply in the multilayer region and suffers high propagation loss due to the optical scattering in the multilayer. Experimentally, typical propagation loss of a Bragg fiber is ~10 dB/m for the light inside the bandgap, and is ~60 dB/m for the light outside the bandgap. The position and width of the fiber bandgap are determined by the refractive indices of the fiber core and Bragg reflector as well as by the thicknesses of the individual layers in the Bragg reflector [6]. Additionally, one end of a Bragg fiber is wet-coated with a thin silver layer serving as a mirror [7]. In our experiment we use a 4 cm-long Bragg fiber with core/cladding diameters to be 240/290 μm. For comparison, we use a commercial 4 cm-long multimode plastic fiber with core/cladding diameters to be 240/250 μm, one of the fiber ends is coated with a silver layer.

grating-based spectrometer are only required to study dependence of the bending sensor sensitivity on the wavelength of operation. Normally, they should be replaced with a laser diode and a power detector. Finally, a glass plate fixed on the nanopositioning stage is used to displace the fiber with increments of 50 μm to provide precisely measured displacements. The glass plate is positioned ~4 mm from the V-groove.

In Fig. 2, we show a typical micrograph of a bent Bragg fiber. In order to extract the bend curvature, the fiber shape is fitted with a circular arc (3-point fitting). For larger displacements, the fiber shape starts to deviate from a simple circular arc, and, as a consequence, error in the determination of the curvature increases with displacement. To estimate the error of the circular arc fitting, we perform a series of 3-point fittings by choosing different sets of fitting points on the fiber, and then find the maximum and minimum of the bending curvatures obtained. The reflection spectra of the Bragg fiber and the regular multimode fiber are presented in Fig. 3 for various values of the curvature.

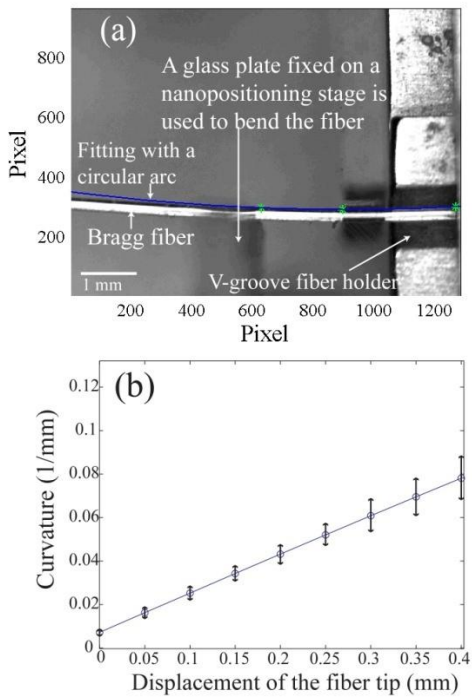

Fig. 2. (a) Fitting of the Bragg fiber shape with a circular arc; (b) curvature of the Bragg fiber as a function of the displacements of the nanopositioning stage. The errors are estimated from the results of several alternative fittings.

Schematic of the Bragg fiber bending sensor is presented in Fig. 1 (b). The light from a superconitnuum source first passes through the beam splitter and it is then coupled into the fiber by an objective. The light coupled into the fiber travels back and forth, as it is reflected back by the silver mirror at the fiber end. The reflected light passes through an objective and is finally redirected by the beam splitter into a spectrometer for analysis. We, note that a supercontinuum source and a

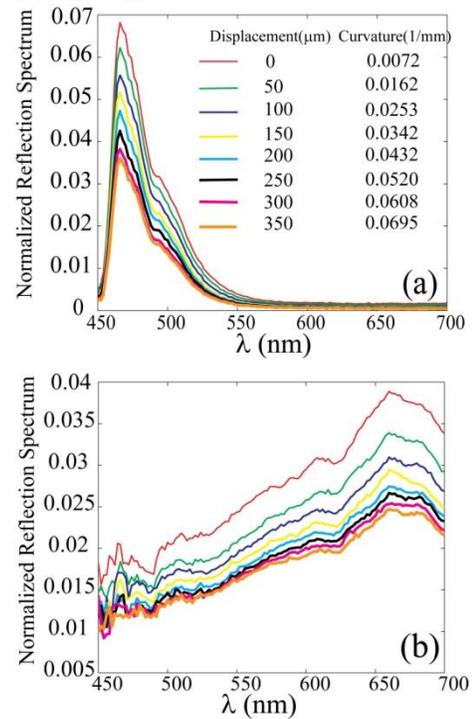

Fig. 3. (a) Reflection spectra of the Bragg fiber and (b) reflection spectra of the regular multimode fiber for various values of the fiber curvature. The displacements of each fiber and the corresponding curvatures of the fitting arc are shown in the inset of (a).

From Fig. 3 it is clear that for both fibers, intensity of the reflected light decreases strongly even for relatively large values of the bending radii 2 -10 cm. For the Bragg fiber, the bandgap center (the frequency of maximal transmission) remains fixed at ~470 nm, while the bandgap itself extends from 460 nm to 520 nm.

From the results shown in Fig. 3, we can now find frequency dependence of the amplitude sensitivity of the two bending sensors. Thus, in Fig. 4 (a, b) we plot the relative intensity difference ($\Delta I/I$) caused by a 50 μm

April 16, 2013

displacement and a 200 μm displacement of the fiber tip. We note that amplitude sensitivity of the regular multimode fiber sensor is approximately constant in the whole spectral range, while amplitude sensitivity of the Bragg fiber increases significantly when operating outside of the fiber bandgap, as predicted in the introductory section. In fact, at the edge or outside of the fiber bandgap, Bragg fiber-based sensor becomes more sensitive than the regular multimode fiber–based sensor. As explained earlier, at these wavelengths, the multilayer reflector of the Bragg fiber acts as an efficient mode stripper which allows scattering of the higher order cladding modes excited by the bend. Unfortunately, we also have to note that outside of the bandgap, propagation loss of the Bragg fibers becomes very high, which leads to a significant signal-to-noise degradation for wavelength longer than 550nm (see Fig. 4(a, b)). Therefore, when choosing the most appropriate sensing wavelength at the bandgap edge, one faces a trade-off between the enhanced sensitivity and decreased signal-to-noise ratio. Besides, comparison between Fig. 4(a) and Fig. 4(b) also suggests that Bragg fiber sensor becomes more sensitive than the regular multimode fiber sensor regardless of operation wavelength, when dealing with large displacements. Thus, in Fig. 4(c), we present the relative change in the total transmitted intensity through the fiber as a function of the fiber displacement. Particularly, when the fiber tip is displaced by $d$, the relative change in the total fiber intensity $\delta$ is defined as:

$$\delta(d) = \sum_\lambda \left[(I(\lambda, 0) - I(\lambda, d)\right] \bigg/ \sum_\lambda I(\lambda, 0) , \qquad (1)$$

where $I(\lambda, 0)$ refers to the wavelength-sensitive output intensity of the fiber with no bending; $I(\lambda, d)$ is the output intensity of the fiber displaced by $d$; $\lambda$ is wavelength. From Fig. 4(c), it is apparent that when the displacement of the fiber is larger than 150 μm, the Bragg fiber sensor becomes more sensitive than the multimode fiber sensor.

In order to estimate amplitude sensitivity, $S$, of the Bragg fiber-based sensor, we use the following definition:
$$S(\lambda) = \left(\partial I(\lambda, \sigma) \big/ \partial \sigma \big|_{\sigma=0}\right) \big/ I(\lambda, 0) ,$$ where $\sigma$ is displacement (or the curvature for bending sensing) [8]. Sensitivity of the Bragg fiber-based sensor at the bandgap edge (550 nm) is ~18.4 (1/mm)$^{-1}$ for bending curvature sensing and is ~3.3 mm$^{-1}$ for displacement sensing, while the corresponding sensitivities of the multimode fiber are ~15.5 (1/mm)$^{-1}$ and ~2.8 mm$^{-1}$, respectively. Assuming that 1% change in the amplitude of reflected light can be detected reliably, the bending and displacement detection limits of the Bragg fiber sensor are ~5×10$^{-4}$ (1/mm) and ~3 μm, respectively.

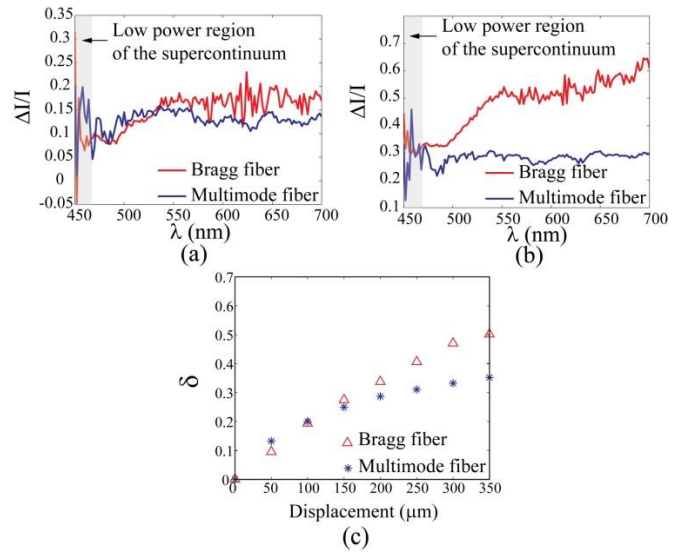

Fig. 4. Relative intensity difference, $\Delta I/I$, of the two Bragg fibers resulted from a (a) 50-μm-displacement and a (b) 200-μm-displacement; (c) Relative change in total intensity, $\delta$, of the fibers resulted from various displacements.

Currently, we are working with SENSORICA group to integrate Bragg fibers into their MOSQUITO bending sensor platform [5]. In this configuration, the Bragg fiber is integrated with a small LED source, a micro-beam-splitter and a micro-photodiode detector, to result in a highly compact bending sensor.

In conclusion, we demonstrate a bending / displacement fiber-optic sensor based on the all-polymer photonic bandgap Bragg fibers. We find that when operated at the edge of a fiber bandgap, Bragg fiber-based sensor is more sensitive to bending as compared to the sensor that uses a regular multimode fiber of comparable diameter. This is due to action of an efficient mode stripper that is present in the Bragg fiber in the form of a multilayer reflector that is present in the Bragg fiber. Displacements as small as 3 μm and bending radii as large as 1m can be detected with the Bragg fiber-based sensor.